\documentclass[aps,showpacs,twocolumn,floatfix,bibnotes]{revtex4}

\usepackage{here}

\usepackage[dvips]{graphicx}

\usepackage[dvips]{graphicx}

\newcommand\pictc[5]{\begin{figure}
                       \centerline{
                       \includegraphics[width=#1\columnwidth, keepaspectratio, height=0.8\textheight]{#3}}
                   \protect\caption{\protect\label{fig:#4} #5}
                    \end{figure}            }
\newcommand\pict[4][1.]{\pictc{#1}{!tb}{#2}{#3}{#4}}
\newcommand\rpict[1]{\ref{fig:#1}}

\newcommand\leqt[1]{\protect\label{eq:#1}}
\newcommand\reqtn[1]{\ref{eq:#1}}
\newcommand\reqt[1]{(\reqtn{#1})}

\newcounter{Fig}

\begin{document}
\begin{sloppy}

\title{Nonlinear surface waves in left-handed materials}

\author{Ilya V. Shadrivov}
\author{Andrey A. Sukhorukov}
\author{Yuri S. Kivshar}
\affiliation{Nonlinear Physics Group, Research School of Physical
Sciences and Engineering, Australian National University,
Canberra ACT 0200, Australia}
\homepage{http://wwwrsphysse.anu.edu.au/nonlinear/}

\author{Alexander A. Zharov}
\affiliation{Institute for Physics of Microstructures, Russian
Academy of Sciences, Nizhny Novgorod 603950, Russia}

\author{Allan D. Boardman and Peter Egan}
\affiliation{Department of Physics, University of Salford, Salford
M5 4WT, United Kingdom}

\begin{abstract}
We study both linear and nonlinear surface waves localized at the
interface separating {\em a left-handed medium} (i.e. the medium
with both negative dielectric permittivity and negative magnetic
permeability) and a conventional (or {\em right-handed})
dielectric medium. We demonstrate that the interface can support
both TE- and TM-polarized surface waves--{\em surface polaritons},
and we study their properties. We describe the intensity-dependent
properties of {\em nonlinear surface waves} in three different
cases, i.e. when both the LH and RH media are nonlinear and when
either of the media is nonlinear. In the case when both media are
nonlinear, we find two types of nonlinear surface waves, one with
the maximum amplitude at the interface, and the other one with two
humps. In the case when one medium is nonlinear, only one type of
surface wave exists, which has the maximum electric field at the
interface, unlike waves in right-handed materials where the
surface-wave maximum is usually shifted into a self-focussing
nonlinear medium. We discus the possibility of tuning the wave group velocity in both the linear and nonlinear cases, and show that group-velocity dispersion,
which leads to pulse broadening, can be balanced by
the nonlinearity of the media, so resulting in soliton propagation.

\end{abstract}

\maketitle

\section{Introduction}

Novel physical effects in dielectric media with both negative
permittivity and negative permeability were first analyzed
theoretically by Veselago \cite{Veselago:1967-517:UFN} who
predicted a number of unusual phenomena including, for example,
negative refraction of waves. Such media are usually known as {\em
left-handed (LH) media} since the electric and magnetic fields
form a left handed set of vectors with the wave vector. The physical
realization of such LH media was demonstrated only recently
\cite{Smith:2000-4184:PRL} for a novel class of engineered
composite materials, now called {\em LH metamaterials}. Such LH
materials have attracted attention not only due to their recent
experimental realization and a number of unusual properties
observed in experiment, but also due to the expanding debates on
the use of a slab of a LH metamaterial as a perfect lens for
focusing both propagating and evanescent waves
\cite{Venema:2002-119:Nature}.

The concept of a perfect lens was first introduced by Pendry
\cite{Pendry:2000-3966:PRL}, who suggested the idea that a slab of
a lossless negative-refraction material can be used for creating a
perfect image of a point source. Although the concept of a perfect
lens is a result of an ideal theoretical model employed in the
analysis~\cite{Pendry:2000-3966:PRL}, the resolution limit of a LH
slab was shown to be independent of the wavelength of the
electromagnetic wave (but can be determined by other factors including losses, spatial dispersion, and others), so that the resolution can be
indeed much better than the resolution of a conventional lens
\cite{Luo:2002-201104:PRB}.

The improved resolution of a LH slab and the corresponding
amplification of evanescent modes even in a lossy LH material, can be
understood from simple physics. Indeed, the near field of an
image, which can not be focused by a normal lens, can be
transferred through the slab of a LH material due to the
excitation of {\em surface waves} (or surface polaritons) at both
interfaces of the slab. Therefore, as the first major step in the
understanding of the amplified transmission of the evanescent
waves, as well as other unusual properties of the LH materials, it
is important to study the properties of different types of surface
wave that can be excited at the interfaces between LH and
conventional (or right-handed, RH) media. Some preliminary studies
in this direction included calculation of the {\em linear}
dispersion properties of modes localized at a single interface or
in a slab of LH material \cite{Ruppin:2000-61:PLA,
Pendry:1998-10096:PRL, Bespyatyh:2001-2043:FTT,
Haldane:cond-mat/0206420, Shadrivov:2003-057602:PRE}.

In this paper, we present a comprehensive study of the properties
of both {\em linear and nonlinear surface waves} at the interface
between semi-infinite materials of two types, left- and
right-handed ones, and demonstrate a number of unique features of
surface waves in LH materials.  In particular, we show the
existence of surface waves of both TE and TM polarizations, a
specific feature of the RH/LH interfaces. We study in detail TE-polarized nonlinear surface waves and suggest an efficient way for engineering the group velocity of surface waves using the nonlinearity of the media. The
dispersion broadening of the pulse can be compensated by the
nonlinearity, thus leading to the formation of surface-polariton
solitons at the RH/LH interfaces with a distinctive vortex-like structure of the energy flow. We must note here, that the presented study is based on the effective medium approximation, which treats the LH materials as homogeneous and isotropic. It can be applied to the manufactured metamaterials, which possess negative dielectric permittivity and negative permeability in the microwave frequency range, when the characteristic scale of the variation of the electromagnetic field (e.g. a field decay length and a wavelength of radiation) is much higher, than the period of the metamaterial. The possibility of preparing isotropic LH materials was studied in \cite{Shelby:2001-489:APL}, where the isotropy of the composite in 2D has been shown. To obtain a negative-refraction material in optics it is suggested that metallic nanowires are used Ref.~\cite{Podolskiy:2002-65:JNOM}. Also, we note that losses is an intrinsic feature of LHM. However, the study of the effect of losses on the guided waves is not in the scope of the present paper.

The paper is organized as follows. In Sec.~\ref{SurfaceWaves} we
study the properties of surface waves in the linear regime. We
consider the most general case of an interface between linear RH
and LH media, and present a classification of TE- and TM-polarized
surface waves localized at the interface. Section~\ref{Nonlinear}
is devoted to the study of the structure and general properties of
nonlinear surface waves. We describe the intensity-dependent
properties of TE-polarized surface waves in three possible cases. In the first
case, we assume that both the LH and RH media are nonlinear. In the
second case, the LH medium is nonlinear, but the RH medium is
assumed to be linear. In the third case, the RH medium is
considered to be nonlinear while the LH medium remains linear. In
all these cases, we take the nonlinear medium have an
intensity-dependent Kerr-like dielectric permittivity. In
Section~\ref{GVE}, we study the frequency dispersion of nonlinear
surface waves. In particular, we demonstrate that the group
velocity of surface waves can effectively be engineered by using
the intensity-dependent dispersion. A detailed analysis is carried
out for the example of nonlinear RH and linear LH media. Finally,
in section~\ref{soliton} we describe the properties of
nonlinear localized modes propagating along the interface, and
predict the existence of surface polariton solitons.

\section{\label{SurfaceWaves} Linear surface waves}

\subsection{Model}

{\em Linear surface waves} are known to exist, under certain
special conditions, at an interface separating two different isotropic
dielectric media. In particular, the existence of TM-polarized
surface waves requires that the dielectric constants of two
dielectric materials separated by an interface have different
signs, whilst for TE-polarized waves the magnetic permeability of
the materials should be of different signs (see, e.g.
Ref.~\cite{Nkoma:1974-3547:JPC, Boardman:EMSurfaceModes} and references therein). Materials with negative $\epsilon$ are readily available (e.g.,
metals excited below a critical frequency), whilst materials
with negative $\mu$ were not known until recently. This explains
why only TM-polarized surface waves have been of interest over the
last few decades.

In this paper, we consider an interface between the RH (medium 1)
and LH (medium 2) semi-infinite media, as shown in the inset of
Fig.~\rpict{XYplane}. The propagation of monochromatic waves with
the frequency $\omega$ is governed by the scalar wave equation,
which for the case of the TE waves is written for the {\em
y}-component of the electric field,
\begin{equation} \leqt{TE-eq}
   \left[ \frac{\partial^2}{\partial z^2}
          + \frac{\partial^2}{\partial x^2}
          + \frac{\omega^2}{c^2} \epsilon(x) \mu(x)
          - \frac{1}{\mu(x)} \frac{\partial \mu(x)}{\partial x}
            \frac{\partial}{\partial x} \right] E_y
   = 0.
\end{equation}
In the case of the TM waves, the scalar wave equations is written
for the {\em y}-component of magnetic field,
\begin{equation} \leqt{TM-eq}
   \left[ \frac{\partial^2}{\partial z^2}
          + \frac{\partial^2}{\partial x^2}
          + \frac{\omega^2}{c^2} \epsilon(x) \mu(x)
          - \frac{1}{\epsilon(x)} \frac{\partial \epsilon(x)}{\partial x}
            \frac{\partial}{\partial x} \right] H_y
   = 0.
\end{equation}
In Eqs.~\reqt{TE-eq} and \reqt{TM-eq}, the functions $\epsilon(x)$
and $\mu(x)$ are dielectric permittivity and magnetic permeability
in a bulk medium, respectively; $\omega$ is the angular wave
frequency, and $c$ is the speed of light in vacuum. The nonzero
components of the magnetic field and of the electric field are found from
the Maxwell's equations, i.e. for TE waves
\begin{equation} \leqt{TE-H-components}
H_z = -\frac{i c}{\omega \mu} \frac{\partial E_y}{\partial x} ; \;\;
H_x =  \frac{i c}{\omega \mu} \frac{\partial E_y}{\partial z},
\end{equation}
and for TM waves
\begin{equation} \leqt{TM-E-components}
E_z =  \frac{i c}{\omega \epsilon} \frac{\partial H_y}{\partial x}
; \;\; E_x = -\frac{i c}{\omega \epsilon} \frac{\partial
H_y}{\partial z},
\end{equation}
respectively. 
Solutions of Eqs.~\reqt{TE-eq} and \reqt{TM-eq} in each linear medium for localized modes, i.e. those propagating along the interface and decaying in transverse direction, have the form
\begin{equation} \leqt{solution-1}
(E_y,H_y) = A_0 e^{i h z - \kappa_{1,2} |x|},
\end{equation}
where $A_0$ is the wave amplitude at the interface, $h$ is a propagation constant,
\[
\kappa_{1,2} =
\left[h^2-\epsilon_{1,2}\mu_{1,2}\left(\frac{\omega}{c}\right)^2\right]^{1/2}
\] is the transverse wave number which characterizes the inverse
decay length of the surface wave in the corresponding medium.

It follows from Eqs.~\reqt{TE-eq} and \reqt{TM-eq} that the tangential components of the electric and magnetic fields change continuously at the interface between two media. These conditions give the dispersion relations for surface waves~\cite{Ruppin:2000-61:PLA},
\begin{equation}
\leqt{TE_lin_dispersion}
\frac{\kappa_1}{\mu_1}+\frac{\kappa_2}{\mu_2}=0,
\end{equation}
and
\begin{equation}
\leqt{TM_lin_dispersion}
\frac{\kappa_1}{\epsilon_1}+\frac{\kappa_2}{\epsilon_2}=0,
\end{equation}
for the cases of the TE- and TM-polarized surface waves,
respectively.

\subsection{\label{linear_properties} Properties of surface waves}

For the analysis presented below, it is convenient to rewrite the
dispersion relations \reqt{TE_lin_dispersion} and \reqt{TM_lin_dispersion} in the
following form,
\begin{equation} \leqt{TE_lin_dispersion1}
h^2=\epsilon_1 \mu_1 \left(\frac{\omega}{c} \right)^2
\frac{Y(Y-X)}{(Y^2-1)},
\end{equation}
and
\begin{equation} \leqt{TM_lin_dispersion1}
h^2=\epsilon_1 \mu_1 \left(\frac{\omega}{c}\right)^2
\frac{X(X-Y)}{(X^2-1)},
\end{equation}
respectively, where we introduced the dimensionless normalized
ratios $X=|\epsilon_2|/\epsilon_1$ and $Y=|\mu_2|/\mu_1$ which
characterize the relative properties of the media creating the
interface. The existence regions for surface waves can be determined from the condition of surface wave localization, i.e. when the transverse wave numbers
$\kappa_{1,2}$ are real, $h>\max{\{\omega \epsilon_1 \mu_1 /c,\;
\omega \epsilon_2 \mu_2 /c}\}$. Existence regions for both
polarizations of surface wave are presented on the parameter plane
$(X,Y)$ in Fig.~\rpict{XYplane}. Along with the polarization, we
determine the type of the wave as {\em forward} or {\em backward},
as discussed below in Sec.~\ref{sec_flow}. We note that
there exist no regions where both TE- and TM-polarized waves
co-exist simultaneously, but both types of surface wave can be
supported by the same interface for different parameters, e.g. for
different frequencies.

\pict{fig01.eps}{XYplane}{Existence regions of surface waves on the
parameter plane $(X, Y$), where $X=|\epsilon_2|/\epsilon_1$ and
$Y=|\mu_2|/\mu_1$. The inset shows the problem geometry.}

\pict{fig02.eps}{freq_dispersion}{Dispersion curves of the
TE-polarized surface waves, for different values of $\epsilon_1$,
shown for the normalized values $\bar{\omega} = \omega/\omega_p$
are $\bar{h} = h c /\omega_p$. Dotted curves marks the dependence
$ \bar{h} = \bar{\omega} \sqrt{ \epsilon_2 \mu_2}$. Dashed line is
the critical frequency  $\omega_1$.}

One of the distinctive properties of the LH materials which has
been demonstrated experimentally is their specific frequency
dispersion. To study the dispersion of the corresponding surface
waves, it is necessary to select a particular form of the
frequency dependence of the dielectric permittivity and magnetic
permeability of the LH medium. A negative dielectric permittivity
is selected in the form of the commonly used function for plasmon
investigations~\cite{Boardman:EMSurfaceModes} and a negative
permeability is constructed in an analogous form (see, e.g.
Ref.~\cite{Ruppin:2000-61:PLA}), i.e.
\begin{equation}
\leqt{freq_dispersion}
    \epsilon_2(\omega) = 1 - \frac{\omega_p^2}{\omega^2} , \;\;\;\;\;
    \mu_2(\omega) = 1 - \frac{F\omega^2}{\omega^2-\omega_r^2} ,
\end{equation}
where losses are neglected, and the values of the parameters
$\omega_p$, $\omega_r$, and $F$ are chosen to fit approximately to the experimental data \cite{Smith:2000-4184:PRL}: $\omega_p/2\pi = 10$~GHz, $\omega_r/2\pi = 4$~GHz, and $F = 0.56$. For this set of parameters, the region in which permittivity and permeability are simultaneously negative is from 4 GHz to 6 GHz.

The dispersion curves of the TE-polarized surface wave (or surface
polariton) calculated with the help of Eq.~\reqt{freq_dispersion}
are depicted in Fig.~\rpict{freq_dispersion} on the plane of the
normalized parameters $\bar{\omega} = \omega/\omega_p$ and
$\bar{h} = h c /\omega_p$. We note that the structure of the
dispersion curves for surface waves depends on the relation
between the values of the dielectric permittivities of the two
media at the characteristic frequency, $\omega_1$, at which the
absolute values of magnetic permeabilities of two media coincide,
$\mu_1=|\mu_2(\omega_1)|$. The corresponding curve in
Fig.~\rpict{freq_dispersion} is monotonically decreasing for
$\epsilon_1>|\epsilon_2(\omega_1)|$, but it is monotonically
increasing otherwise, i.e. for
$\epsilon_1<|\epsilon_2(\omega_1)|$. Only the
first case was identified in the previous analysis reported in
Ref.~\cite{Ruppin:2000-61:PLA}. The change of the slope of the
curve (the slope of the dispersion curve represents the group
velocity) with the variation of the dielectric permittivity of the
RH medium can be used for group velocity engineering, which we
discuss in Sec.~\ref{GVE}.

The critical value of dielectric permittivity
$|\epsilon_2(\omega_1)|$ for the case of a nonmagnetic RH medium
($ \mu_1 = 1 $) is found from the dispersion
relations~\reqt{TE_lin_dispersion} and \reqt{freq_dispersion}, and
it has the form
\begin{equation} \leqt{critical_epsilon}
    \epsilon_c = |\epsilon_2(\omega_1)| = \left(1-\frac{F}{2} \right)  \left(\frac{\omega_p}{\omega_r}\right)^2-1
\end{equation}
For the parameters specified above, this critical value is
$\epsilon_c = 3.5$.

The change of the dispersion curve from monotonically increasing
to monotonically decreasing, shown in
Fig.~\rpict{freq_dispersion}, is connected with a change in the
direction of the total power flow in the wave, as discussed below.

\subsection{\label{sec_flow} Energy flow near the interface}

The energy flow is described by the Poynting vector, which defines
the energy density flux averaged over the period $T=2\pi/\omega$,
and can be written in the form
\begin{equation} \leqt{PoyntingVector}
{\bf S}=\frac{c}{8\pi}\rm{Re}\left[{\bf E}\times {\bf H}^*\right],
\end{equation}
where ${\bf E}, {\bf H}$ are the complex envelopes of the electric
field and magnetic field of a surface wave, respectively, and the
asterisk stands for the complex conjugation.

A uniform surface wave propagating along the interface has
only one non-zero component of the averaged Poynting vector, $|{\bf
S}|=|S_z|$. The energy flux in the RH and LH media is an integral
of the Poynting vector over the corresponding semi-infinite
spatial region,
\begin{equation}\leqt{power_stat1}
P_1 = \int_{-\infty}^{0} S_z \, dz = \frac{B h}{\kappa_1}\left\{
\begin{array}{rcl}
1/\mu_1; \; \mbox{for TE},\\
1/\epsilon_1; \; \mbox{for TM},
\end{array}
\right.
\end{equation}
\begin{equation}\leqt{power_stat2}
P_2 = \int_{0}^{\infty} S_z \, dz = \frac{B h}{\kappa_2}\left\{
\begin{array}{rcl}
1 /\mu_2; \; \mbox{for TE},\\
1/\epsilon_2; \; \mbox{for TM},
\end{array}
\right.
\end{equation}
where the constant $B = c^2 A_0^2/16\pi\omega$. We
note that the electromagnetic energy flow is in opposite
directions at either side of the interface, as was also predicted in ~\cite{Nkoma:1974-3547:JPC} for TM polarized waves. The total energy flux
in the forward $z$-direction is defined as the sum, $P=P_1+P_2$, and it is
found as
\begin{equation}\leqt{power_stat_total}
    P = \left( 1-XY \right)
    \frac{B h \omega^2 \epsilon_1 \mu_1 }{ \kappa_1 \kappa_2 c^2}
	\left\{
    \begin{array}{rcl}
        \left( 1+Y^2 \right) /\left[ Y (\mu_1\kappa_1-\mu_2\kappa_2)
        \right],\\
        \left( 1+X^2 \right) /\left[ X (\epsilon_1\kappa_1-\epsilon_2\kappa_2)
        \right],
   \end{array}
    \right.
\end{equation}
for the TE and TM waves, respectively. The total energy flux is
positive for $XY<1$, and negative for $XY>1$. The surface waves
are forward or backward, respectively.  The corresponding types of
surface waves determined from this analysis are labelled in
Fig.~\rpict{XYplane}.

\section{\label{Nonlinear} Nonlinear surface waves}

\subsection{ Nonlinear LH/RH interface}

Nonlinear surface waves at an interface separating two
conventional dielectric media have been analyzed extensively for
several decades starting from the pioneering
paper~\cite{Litvak:1968-1911:IVR}. In brief, one of the major
findings of those studies is that the TE-polarized surface waves
can exist at the interface separating two RH media provided that
{\em at least one of these is nonlinear}, but that {\em no surface
waves exist in the linear limit}.

In this section, we study TE-polarized nonlinear surface waves assuming that
both media are nonlinear, i.e. they display a Kerr-type
nonlinearity in their dielectric properties, namely
\begin{equation} \leqt{nonlinear_kerr}
\epsilon_{1,2}^{NL} = \epsilon_{1,2} + \alpha_{1,2}|E|^2,
\end{equation}
where the first term characterizes the linear properties, i.e.
those in the limit of vanishing wave amplitude.

First, we should mention that the recent systematic study of
nonlinear properties of metallic composites~\cite{Zharov:2003-037401:PRL}
suggested the possibility of hysteresis-type nonlinear effects in
a structure consisting of arrays of split-ring resonators (SRRs)
and wires embedded in a nonlinear dielectric medium. Such effects
can also be caused by a nonlinear dielectric material placed in
the slits of the SRRs, which results in an intensity-dependent
capacitance of the slit. These hysteresis effects can be avoided
if the structure is filled by a nonlinear dielectric material
except in the SRR slits. In what follows, we consider such
composite structures for which the nonlinear properties can be
characterized by Eq.~\reqt{nonlinear_kerr} valid far from the
resonances.

For a conventional (or right-handed) dielectric medium, positive
$\alpha_1$ corresponds to a self-focusing nonlinear material,
whilst negative $\alpha_1$ characterizes defocusing effects in the
beam propagation. However, this classification becomes reversed in
the case of LH materials and, for example, a self-focusing LH
medium corresponds to negative $\alpha_2$. Indeed, taking into
account relation \reqt{nonlinear_kerr}, we rewrite
Eq.~\reqt{TE-eq} for the case of the TE-polarized wave in
nonlinear media as follows,
\begin{equation} \leqt{Helm}
    \frac{\partial^2 E}{\partial z^2}
    + \frac{\partial^2 E}{\partial x^2} + \left(\frac{\omega}{c}\right)^2
    \left(\epsilon\mu +  \mu\alpha |E|^2 \right) E = 0.
\end{equation}
According to Eq. \reqt{Helm}, the sign of the product $\mu \alpha$ determines the type of nonlinear self-action effects which occur. Therefore, in a LH medium with negative
$\mu_2$ {\em all nonlinear effects are opposite} to those in RH media with positive $\mu_1$, for the same $\alpha$. Below, we
assume for definiteness that both LH and RH materials possess
self-focusing properties, i.e. $\alpha_1>0$ and $\alpha_2<0$.

We look for the stationary solutions of Eq.~\reqt{Helm} in the
form $E_{1,2}(x,z) = \Psi_{1,2}(x) \exp{(i h z)}$. Then, the profiles of the spatially localized wave envelopes $\Psi_{1,2}(x)$ are found as \cite{Zakharov:1972-62:JETP}
\begin{equation} \leqt{Helm_solution}
    \Psi_{1,2}(x) = \left(\eta_{1,2} \sqrt{2/\alpha_{1,2} \mu_{1,2}} \right) \; {\rm sech} [ \eta_{1,2} (x - x_{1,2})],
\end{equation}
where $\eta_{1,2} = \kappa_{1,2} c/\omega$ are normalized
transverse wave numbers, $x_{1,2}$ are centers of the sech-functions which should be chosen to satisfy the continuity of the tangential components of the electric and magnetic fields at the interface. These conditions can be presented in the form of two transcendental equations,
\begin{equation} \leqt{coord1}
    \tanh^2{(\eta_1 x_1)} = \left(
            1 - \frac{\alpha_1 \mu_1 \eta_2^2}{\alpha_2 \mu_2 \eta_1^2}
                    \right)
                            \left(
            1 - \frac{\alpha_1 \mu_2}{\alpha_2 \mu_1}
                            \right)^{-1},
\end{equation}
and
\begin{equation} \leqt{coord2}
\frac{\eta_2}{\mu_2}\tanh{(\eta_2 x_2)} =
    \frac{\eta_1}{\mu_1}\tanh{(\eta_1 x_1)}.
\end{equation}
Note, that if the parameters  $(x_1,x_2)$ correspond to one of the
solutions of the equation, then $(-x_1,-x_2)$ gives another
solution. Two waves described by these solutions have the same
wave number, but they correspond to different transverse
structures of the surface wave, one of which has a maximum of the
intensity at the interface, and the other one that has two humps
shifted into the media.

In a degenerate case, when $(\alpha_1,\mu_1,\epsilon_1) = (\left|
\alpha_2 \right|,  \left| \mu_2 \right|,\left| \epsilon_2
\right|)$, the system has an infinite number of solutions. Indeed,
any $(x,-x)$ pair will describe a stationary solution for the
surface wave. Such waves have a zero total energy flux because of
the symmetry of the solution.

The energy flow in this wave can be written in the form
\begin{equation} \leqt{power_flow_2nl}
P = P_0 \gamma \left[ \frac{\eta_1 \alpha_2 \mu_2}{\alpha_1 \mu_1^2}
              +\frac{\eta_2}{\mu_2}
              +\frac{\eta_1}{\mu_1}
                    \left( 1
                            -\frac{\alpha_2 \mu_2}{\alpha_1 \mu_1}
                    \right) \tanh{(\eta_1 x_1)}
        \right],
\end{equation}
where $P_0=c^2/4\pi\omega\alpha_2\mu_2$, and $\gamma = h c/\omega$
is the normalized wave number. We now consider the surface waves in the non-degenerate case when only $\alpha_1 = \left| \alpha_2 \right|$. The dependence of the normalized energy flux $P/P_0$ on the parameter $\gamma$ is shown in Fig.~\rpict{2nl} for the cases when linear waves are forward or
backward, respectively. Corresponding transverse wave structures
are shown in the insets.

\pict{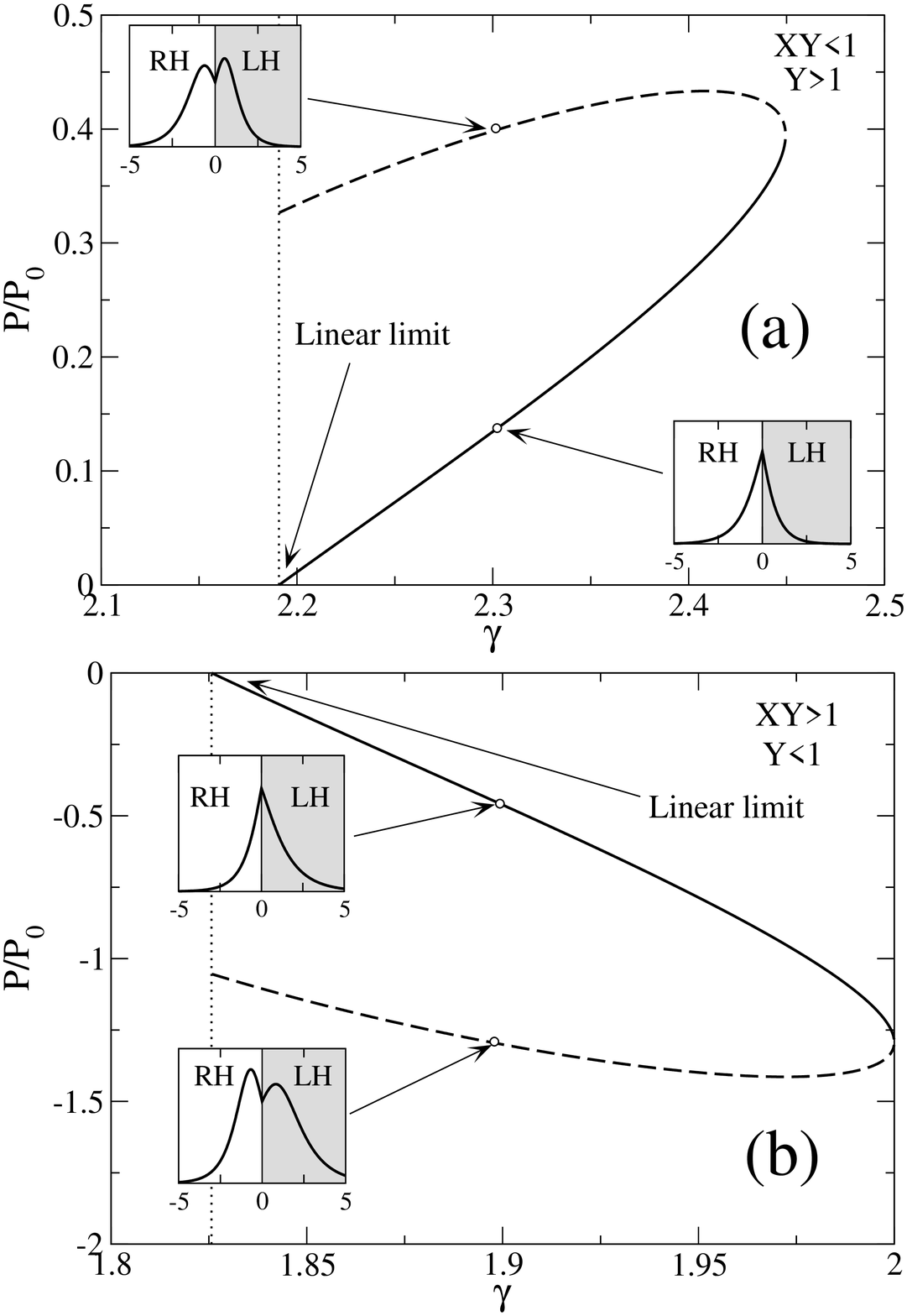}{2nl}{Normalized energy flux {\em vs.} normalized
wavenumber $\gamma = h c/\omega$ for the nonlinear surface waves
in two cases: (a) $XY<1$, $Y>1$, and (b) $XY>1$, $Y<1$. Solid
curve corresponds to a one-humped structure, dashed --
double-humped structure. The insets shows the structure of the
surface waves at the points indicated by arrows. Dotted lines
denote the linear surface wave wavenumber.}

The linear limit corresponds to the case $P \to 0$ when $x_1 \to
+\infty$ and $x_2 \to -\infty$. Moving along the curves, the
centers of the sech-functions move toward the interface and
at the point with $dP/d\gamma = \infty$, $x_1=x_2=0$, and from
that point along the dashed line $x_1 \to -\infty, x_2 \to
+\infty$, thus revealing the two-humped transverse structure of
the surface wave. Note that the forward (backward) wave in the linear
case remains forward (backward) in the nonlinear case, i.e. the type
of the mode is determined by the linear parameters of the system
and can be found using the diagram in Fig.~\rpict{XYplane}.

\subsection{Nonlinear LH/linear RH interface}

Next, we consider surface waves propagating along an interface
between linear RH and nonlinear LH media (see the inset in
Fig.~\rpict{power_dispersion}) having the negative nonlinear
coefficient $\alpha_2$ and, thus, displaying the self-focusing
properties. The transverse structure of the stationary surface
wave has the form:
\begin{equation} \leqt{structure}
    \Psi(x) =\left\{
    \begin{array}{lr}
        E_0 \exp{(\eta_1 x)}, \; x < 0, \\
        \left( 2/\alpha_2 \mu_2\right)^{1/2} \eta_2 \, {\rm sech}
        \left[ \eta_2(x-x_0) \right], \;
        x>0,
    \end{array}
    \right.
\end{equation}
where $E_0$ and $x_0$ are two parameters which should be
determined from the continuity conditions at the interface for the
tangential components of the electric and magnetic fields,
\begin{eqnarray}
\leqt{nonlin_dispersion}
    \tanh{\left(\eta_2 x_0\right)} = \mu_2 \eta_1/(\mu_1 \eta_2), \nonumber\\
    E_0 = (2/\alpha_2 \mu_2)^{1/2} \eta_2 \,
    {\rm sech}(\eta_2 x_0).
\end{eqnarray}
Analyzing the relations \reqt{nonlin_dispersion}, we find that a
surface wave always has the maximum of field intensity at the
interface. This is in a sharp contrast to the nonlinear surface
waves excited at the interface separating two RH media, when the
electric field has maximum shifted into a self-focusing nonlinear
medium~\cite{Boardman:1985-1701:JQEL}.

The corresponding nonlinear dispersion relation of the surface
waves is found in the form
\begin{equation} \leqt{nl_disp}
    \frac{\eta_1}{\mu_1}+\frac{\eta_2}{\mu_2}
    \left(1 - \frac{A_1^2}{\eta_2^2}\right)^{1/2}=0,
\end{equation}
where $A_1=E_0 (\alpha_1\mu_1/2)^{1/2}$ is the normalized electric
field amplitude at the interface. Equation \reqt{nl_disp} reduces
to the linear dispersion relation \reqt{TE_lin_dispersion} in the
small-amplitude limit, i.e. when $A_1 \to 0$.

\pict{fig04.eps}{power_dispersion}{Normalized energy flux
{\em vs} normalized wavenumber $\gamma = h c/\omega$ for the
nonlinear surface waves at the nonlinear LH/Linear RH interface.
Surface waves can be both forward (positive energy flux) and
backward (negative energy flux). The inset shows the geometry of
the problem. The solid line shows the transverse wave profile, dotted line shows the continuation of the solution in nonlinear medium \reqt{structure} to the linear medium, dashed line indicates the position of the center $x_0$ of the sech - function.}

The energy flux $P$ associated with the nonlinear surface wave can
be calculated in the form
\[
\leqt{power}
    P = P_0 \gamma \eta_2
        \left(
        1+\frac{\mu_2\eta_1}{\mu_1\eta_2}
        \right)
        \left[ \frac{2}{\mu_2} +
        \frac{\eta_2}{\eta_1\mu_1}
        \left(
        1-\frac{\mu_2\eta_1}{\mu_1\eta_2}
        \right)
        \right],
\]
where $P_0$ is defined above.

As an example, we consider the case $XY < 1$ for which, as we have
shown above, only forward surface waves can exist at the interface
between two linear media.  However, nonlinear surface waves can be
either forward or backward, as demonstrated in
Fig.~\rpict{power_dispersion}. For $Y<1$, there exists {\em no
linear limit} for the existence of the surface waves, while in
other two regions the results for linear surface waves are
recovered in the limit $P\to 0$. The point on the curve
corresponding to $P=0$ in the case $Y<1$ describes the wave of
finite amplitude in which the energy flows on either side of the
interface are balanced. Such a wave does not exist in the linear
limit.

\subsection{Linear LH/nonlinear RH interface}

Finally, we consider the case when the LH material is linear,
while the RH medium is nonlinear. In such a geometry, the
dispersion relation for the TE-polarized waves  has the form
\begin{equation} \leqt{nonlin_dispers_2}
    \frac{\eta_2}{\mu_2}+\frac{\eta_1}{\mu_1}
    \left(1- \frac{A_2^2}{\eta_1^2}\right)^{1/2}=0,
\end{equation}
where $A_2=E_0 \sqrt{\alpha_2\mu_2/2}$ is the normalized amplitude
of the electric field  at the interface.

The dependence of the normalized energy flux on the wave number of
the surface wave is shown in Fig.~\rpict{power_dispersion2} for
$XY> 1$. In contrast to the linear waves, the nonlinear surface
waves can be either forward or backward (see Fig.~\rpict{XYplane}).
In analogy with the case $Y<1$ for the nonlinear LH/linear RH
interface, it can be shown that there exists no small-amplitude
limit for the nonlinear surface waves for $Y > 1$ . For $XY < 1$
only forward travelling waves exist when $Y > 1$, reproducing the
property of the corresponding linear waves.

\pict{fig05.eps}{power_dispersion2}{Normalized energy flux
{\em vs} normalized wavenumber $\gamma = h c/\omega$ for the
nonlinear surface waves at the linear LH/nonlinear RH interface.
Surface waves can be either forward (positive energy flux) or
backward (negative energy flux). The inset shows the geometry of
the problem.}

\section{\label{GVE} Frequency dispersion of nonlinear surface waves}

We have demonstrated in Sec. \ref{linear_properties} that the frequency dispersion of surface waves depends on the dielectric permittivity of the RH medium (see Fig.~\rpict{freq_dispersion}). These results suggest that the dispersion type can be switched between normal and anomalous if the RH medium is nonlinear.

To demonstrate this property, we study the properties of nonlinear surface waves near the critical point and select $\epsilon_{1}=3.4$, in order to stay just below the critical value $\epsilon_{c}$ corresponding to the linear case. It should be mentioned here that although the negative permeability of the LH composite material is necessary for the
existence of TE surface waves in the present model, it has been
shown~\cite{Maradudin:1981-341:ZPCM} that TE surface waves do
exist at the interface between a RH nonlinear plasma and a RH
nonlinear dielectric medium with a constant zero-field
permittivity, provided that the nonlinear parameter in the plasma
exceeds that in the dielectric medium.  In that case, since $\mu =
1$ in both media, it has also been
shown~\cite{Maradudin:1981-341:ZPCM} that the relation for
wave intensity at the boundary and the frequency are independent
of the wave number.  In the present case, where the LH composite
material has a permeability not equal to unity, this result is no
longer valid. In our problem, Eq.~\reqt{nonlin_dispers_2} provides
a dependence between the three variables $h$, $A_2$ and $\omega$,
so that the wave intensity at the boundary depends on both $h$ and
$\omega$.

As was shown for the case of linear surface waves, a change of the
slope of the dispersion curve takes place at the critical value of
dielectric permittivity of the RH medium~\reqt{critical_epsilon}.
In the nonlinear case, the dielectric permittivity of the RH
medium depends on the field intensity and, in particular, it
exceeds the critical value when the wave amplitude $A_2$ becomes
larger than the threshold value $A_{2c}$ given by the equation
\begin{equation} \leqt{A_crit}
    A_{2c}^2=\left(\frac{\omega_p}{\omega_r}\right)^2 \left(1-\frac{F}{2} \right)
    -\left( 1+\epsilon_1 \right).
\end{equation}
For the media parameters corresponding to Fig.~\rpict{freq_dispersion}, equation~\reqt{A_crit} gives: $A_{2c}=0.3162$.
Figure~\rpict{nonlin_freq_dispersion} shows the dispersion curves
for three different wave amplitudes. As was shown in Sec.~\ref{linear_properties}, the frequency dispersion is normal for the wave amplitudes below the critical value \reqt{A_crit}, and it is negative, otherwise. One can also notice from Fig.~\rpict{nonlin_freq_dispersion} that the
existence region for surface waves depends on the wave
amplitude $A_2$. In Fig.~\rpict{nonlin_h_dispersion} we present
this region of wave existence on the plane of the wave amplitude and
normalized frequency. The existence region of the backward surface
waves below the critical value collapses at $A_{2c}$, and it
expands in the region of the forward surface waves above the
threshold. Note, that the forward and backward waves exist at
different frequencies only.

The existence regions of the surface wave can be explained from
the viewpoint of the physics of wave localization. The wave
localization is determined by the normalized transverse
wavenumbers $\eta_{1,2}$, which define the inverse decay length of the surface wave in the corresponding medium. The conditions $\eta_1 = 0$ and $\eta_2=0$ correspond to the delocalized waves,  and determine the boundary of the existence region. Figure~\rpict{nonlin_a_dispersion} shows the dependence of the normalized wave number on the wave amplitude for different frequencies. These curves represent the horizontal cross-sections of the region of the wave existence shown in Fig.~\rpict{nonlin_h_dispersion}. The dashed line in
Fig.~\rpict{nonlin_a_dispersion} shows the boundary of
localization of the wave in the LH material. Comparing
Fig.~\rpict{nonlin_a_dispersion} and Fig.~\rpict{nonlin_h_dispersion}, we come to the conclusion that in Fig.~\rpict{nonlin_h_dispersion} the upper boundary for the existence of the forward waves (below $A_{2c}$) and the lower
boundary for the backward wave existence (above $A_{2c}$) are
determined by the wave localization in the LH material.

\pict{fig06.eps}{nonlin_freq_dispersion}{Normalized
frequency vs. normalized wave number for different values of the
amplitude  $A_2$. Dashed line corresponds to the line $\omega_1$
in Fig.~\rpict{freq_dispersion}; dotted line is $\eta_2=0$.}

\pict{fig07.eps}{nonlin_h_dispersion}{Existence region
of nonlinear surface waves (shaded). The curves show the amplitude
$A_2$ vs. normalized frequency, for different values of the
normalized wave number (marked at the curves). All curves
intersect at the critical point $A_2=A_{2c}$.}

\pict{fig08.eps}{nonlin_a_dispersion}{Normalized wave
number vs. normalized field amplitude $A_2$, for different
frequencies. Dashed line is $\eta_2=0$. The curves on the left of
$A_{2c}$ meet the wave number axis at the values corresponding to
the linear case. There exist no linear solutions to the right of
the critical vertical line $A_2=A_{2c}$}

The power flow in the linear LH composite medium is given by the
result
\begin{equation} \leqt{power2}
P_2=\frac{c^2 \gamma}{16\pi\omega\mu_2\eta_2} E_0^2,
\end{equation}
and, using the boundary value
technique~\cite{Boardman:1985-1701:JQEL}, the power flow in the
nonlinear half-space (RH medium) can be obtained in the following
form
\begin{equation} \leqt{power1}
P_1=\frac{c^2 \gamma }{4\pi\omega\mu_1^2\alpha_1}
                \left(
    \eta_1-\sqrt{\eta_1^2-E_0^2\frac{\alpha_1\mu_1}{2}}
                \right).
\end{equation}
We note here that, although the nonlinear coefficient $\alpha_1$
appears as a factor in the denominator of Eq.~\reqt{power1}, the
reduction to the linear case (when $\alpha_1\to 0$) can be
performed in a straightforward way by expanding Eq.~\reqt{power1}
as for small $\alpha_1$ as follows,
\begin{equation} \leqt{power1_2}
    P_1 = \frac{c^2 \gamma}{16 \pi \omega \mu_1 \eta_1} E_0^2 + O(\alpha_1^2),
\end{equation}
which has the same form as Eq.~\reqt{power2}.

The absolute value of the ratio of the power flow in the RH
nonlinear medium to the power flow in the LH composite medium is
depicted in Fig.~\rpict{power_ratio}.  Above the critical value
$A_c$, there exists a value of the wave number at which the
power flow is positive.  In this region, there exist a forward
travelling surface wave.  Note that no matter what the value of
the field intensity at the boundary is, there always exists some
value of the wave number where the flow is negative and there
exists a backward travelling surface wave.  This can be seen by
reference to Eq.~\reqt{power2}.  The presence of $\eta_2$ in the
denominator of Eq.~\reqt{power2} means that as $h$ approaches a
value that makes $\eta_2 = 0$ (i.e. a dispersion curve in
Fig.~\rpict{nonlin_freq_dispersion} approaches the dashed line)
the negative power flow in the LH material dominates the total
power flow.  Conversely, as $\eta_2$ increases from zero, the
negative power flow in the LH medium decreases so that, provided
the intensity of the electric field at the boundary is high
enough, the positive power flow in the nonlinear dielectric
dominates.

\pict{fig09.eps}{power_ratio}{The absolute value of the ratio of
the power flow in the nonlinear dielectric to that in the LH
material. Above the critical value $A_c$, the power flow in the
nonlinear half-space dominates giving a forward travelling wave.
As $\kappa_2$ approaches zero towards the left-hand side of the
curves, the power in the left-handed medium dominates.}

\section{\label{soliton} Nonlinear pulse propagation and surface-wave solitons}

\subsection{Envelope equation}

Propagation of pulses along the interface between RH and LH media
is of a particular interest, since it was shown before \cite{Nkoma:1974-3547:JPC} for TM modes that the energy fluxes are {\em directed oppositely} at either side of the interface. Therefore, we can expect that the energy flow in a pulse with finite temporal and spatial dimension has a nontrivial form \cite{Shadrivov:2003-057602:PRE} and, in particular, it can be associated with a vortex-like structure of the energy flow.

We analyze the structure of surface waves of both temporal and
spatial finite extent that can exist in such a geometry. To
obtain the equation describing the pulse propagation along the
interface, we look for the structure of a broad electromagnetic
pulse with carrier frequency $\omega_0$ described by an
asymptotic multi-scale expansion with the main terms of the
general form
\begin{eqnarray}
\leqt{formEq}
    \Psi =  e^{ih_0z-i\omega_0 t}\left[
        \Psi_0(x) A(\xi,t) - \right.
        \nonumber\\
       \left. i \Psi_1(x) \frac{\partial A(\xi,t)}{\partial \xi} +
        \Psi_2(x,\xi,t) + \ldots \right] ,
\end{eqnarray}
where the field $\Psi$ stands for the components $(E_y, H_x, H_z)$
of a TE-polarized wave, the first term $\Psi_0=(E_{y0}, H_{x0},
H_{z0})$ describes the structure of the mode at the carrier
frequency $\omega_0$, $\Psi_1$ is the first-order term of the
asymptotic series which can be found as
$\Psi_1=\partial\Psi_0/\partial h$, and $\Psi_2$ is the
second-order term. Here $A$ is the pulse envelope, $h_0$ is the
wave number corresponding to the carrier frequency $\omega_0$,
$\xi=z-v_g t$ is the pulse coordinate in the reference frame
moving with the group velocity $v_g = \partial \omega/\partial h $.
Substituting Eq.~\reqt{formEq} into Eq. \reqt{Helm} and using
the Fredholm alternative theorem \cite{Korn:MathHandbook}, one can obtain the equation for the evolution of the field envelope
\begin{equation} \leqt{NLS}
    i\frac{\partial A}{\partial t}
    +\frac{\delta}{2} \frac{\partial^2 A}{\partial \xi^2} -
    \omega_2(h)|A^2| A =0,
\end{equation}
where the coefficient $\delta=\partial^2\omega/\partial h^2$
stands for the group-velocity dispersion (GVD) which determines the pulse broadening and can be calculated from the dispersion relations, $\omega_2(h)=\left(\partial\omega_{NL}/\partial A^2 \right) |_{A=0}$ is the effective nonlinear coefficient calculated with the help of the nonlinear dispersion relation. The NLS equation has a solution in the form of a bright soliton localized at the interface, provided the GVD ($\delta$) has the opposite sign to the sign of the nonlinear coefficient ($\omega_2$) (see, e.g., \cite{Kivshar:OpticalSolitons} and references therein). The existence of the surface polariton solitons has been predicted in a number of structures supporting nonlinear guided waves (see, e.g., Ref.~\cite{Boardman:1986-8273:PRB} and references therein). 

The effective nonlinear coefficient for the case of an interface between the nonlinear LH medium and the
linear RH medium, can be found using Eq.~\reqt{nonlin_dispersion},
\begin{equation} \leqt{nonlin_coefficient}
    \omega_2(h)=-\frac{\alpha \mu_1 \kappa_1 \kappa_2 \omega^2 }{
    4 h c^2(\epsilon_2\mu_2-\epsilon_1\mu_1)}
    \frac{d\omega}{dh}.
\end{equation}
The signs of the group velocity $d\omega/dh$ and of the parameter
$\delta$ can be determined from the Fig.~\rpict{freq_dispersion}.
As a result, for any reasonable values of dielectric permittivity
and magnetic permeability of the RH medium, there exists a range
of frequencies for which $\omega_2 \cdot \delta <0$, indicating
the possibility of exciting surface polariton solitons.

To study the energy flow in such a surface-polariton soliton, we
use the asymptotic expansions~\reqt{formEq} for the field components,
and from Eq.~\reqt{PoyntingVector} we obtain the energy flow structure described by their components
\begin{equation} \leqt{Poynting_temporal_z}
S_z = \frac{c^2 h_0}{8\pi\omega_0 \mu} E_0^2 \left| A \right|^2,
\end{equation}
\begin{eqnarray} \leqt{Poynting_temporal_x}
S_x = \frac{c^2}{8\pi\omega_0\mu} \left[
    \frac{\partial E_0}{\partial h} \frac{\partial E_0}{\partial x} -
    E_0 \left( \frac{\partial^2 E_0}{\partial h \partial x} -
    \frac{v_{gr}}{\omega_0}\frac{\partial E_0}{\partial x} \right) \right] \cdot
	\nonumber\\
    {\rm Re}\left( A \frac{\partial A^*}{\partial \xi} \right),
\end{eqnarray}
The structure of the Poynting vector in the surface-wave pulse is
shown in Fig. (\rpict{vortex}),  where it is clearly seen that the
energy rotates in the localized region creating a vortex-type
energy distribution in the wave. The difference between the
Poynting vector $|S_z|$ integrated over the RH medium and that
calculated for the LH medium determines the resulting group
velocity of the surface wave packet.

\pict{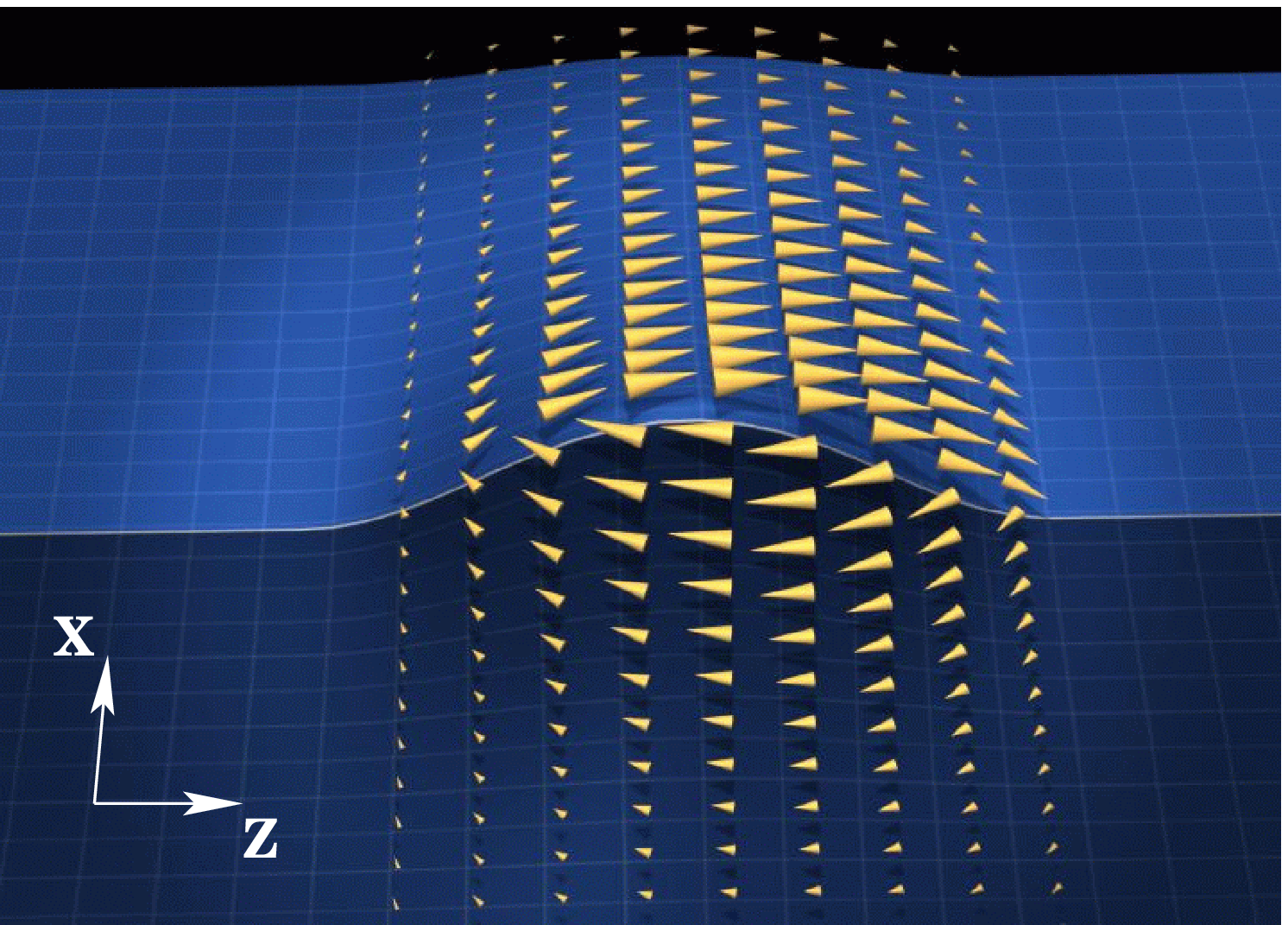}{vortex}{A vortex-like distribution of the Poynting
vector in a surface wave soliton propagating along the LH/RH
interface.}

We note the distinctive vortex-like structure of the surface waves at the interface separating RH and LH media follows as a result of the opposite signs of the dielectric permittivity for the TM-polarized waves and the magnetic permittivity for the TE-polarized waves. These conditions coincide with the conditions for the existence of the corresponding surface waves and, therefore, surface polaritons should always have such a distinctive vortex-like structure.

\section{\label{Conclusions} Conclusions}

We have presented a systematic study of both linear and nonlinear
surface waves supported by an interface between a left-handed
metamaterial and a conventional dielectric medium. For the linear
regime, we have extended some earlier theoretical results and
analyzed different types of surface waves and their existence
regions. In particular, we have demonstrated that the structure of
the energy flow in a spatially localized wave-packet of the
surface waves propagating along the interface has a vortex-like
structure. For the case of nonlinear surface waves, we have
demonstrated that when only one of the media is nonlinear the
maximum amplitude of the spatially localized wave does not shift
out the interface. This result is in a sharp contrast to the case
of an interface between linear and nonlinear right-handed
materials where the maximum of a surface wave is always shifted
into a self-focusing nonlinear medium. We have demonstrated that
the group velocity of nonlinear surface waves can be controlled by
changing the intensity of the electromagnetic field and, in
particular, the surface wave can be switched from the forward
propagating one to the backward propagating one by varying the
field intensity only. In addition, we have obtained the conditions
for the existence of the surface-polariton solitons at the
metamaterial interface.

\end{sloppy}
\end{document}